# Spin states of zigzag-edged Möbius graphene nanoribbons from first principles

*De-en Jiang[1,*] and Sheng Dai[1,2]*

[1]Chemical Sciences Division and [2]Center for Nanophase Materials Sciences, Oak Ridge National Laboratory, Oak Ridge, Tennessee 37831

jiangd@ornl.gov

*To whom correspondence should be addressed. E-mail: jiangd@ornl.gov. Phone: (865)574-5199. Fax: (865) 576-5235.

Möbius graphene nanoribbons have only one edge topologically. How the magnetic structures, previously associated with the two edges of zigzag-edged flat nanoribbons or cyclic nanorings, would change for their Möbius counterparts is an intriguing question. Using spin-polarized density functional theory, we shed light on this question. We examine spin states of zigzag-edged Möbius graphene nanoribbons with different widths and lengths. We find a triplet ground state for a Möbius cyclacene, while the corresponding two-edged cyclacene has an open-shell singlet ground state. For wider zigzag-edged Möbius graphene nanoribbons, the total magnetization of the ground state is found to increase with the ribbon length. For example, a quintet ground state is found for a zigzag-edged Möbius graphene nanoribbon. Local magnetic moments on the edge carbon atoms form domains of majority and minor spins along the edge. Spins at the domain boundaries are found to be frustrated. Our findings show that the Möbius topology (that is, only one edge) causes zigzag-edged Möbius graphene nanoribbons to favor one spin over the other, leading to a ground state with non-zero total magnetization.

Keywords: Möbius ribbon, electronic structure, spin state, density functional theory, zigzag edge

## 1. Introduction

Möbius strips are fascinating not only in topology but also in chemistry. When the mathematically abstracted one-sided surface dresses as molecules, it surely enticed and challenged chemists' minds. Because aromatic hydrocarbons tend to have flat geometry due to the π-conjugation, it is more challenging to synthesize Möbius aromatic hydrocarbons but more interesting to examine their electronic structures.[1,2] The concept of Möbius aromaticity and delocalization has been widely discussed in the literature.[1-7] Moreover, the first stable π-system with one Möbius twist (a Möbius annulene) has been synthesized[8,9] and encouraging efforts towards synthesizing Möbius aromatic hydrocarbons have been reported recently.[10]

Zigzag-edged graphene nanoribbons as a special kind of π-system have received a lot of interest in the past several years.[11-15] Due to the zigzag boundaries and the bipartite lattice structure of zigzag-edged graphene nanoribbons, the π-electrons localize at the zigzag edges[16] and form spin orders, leading to an antiferromagnetic ground state with one edge spin up and the other spin down.[17] This antiferromagnetic ground state has been shown to be the ground state of not only infinitely long zigzag-edged graphene nanoribbons[18] but also finite-sized nanographenes,[19] higher acenes,[20] and cyclacenes and short zigzag nanotubes.[21]

An interesting question arises as to what would happen to the spin order and the antiferromagnetic state when a zigzag-edged graphene nanoribbon transforms into a zigzag-edged Möbius graphene nanoribbon, causing the two zigzag edges to become one (Figure 1). Using the Hubbard model with the unrestricted Hartree-Fock approximation, Wakabayashi and Harigaya[22] found that the total magnetization of the Möbius strip is zero for even N (ribbon width) and non-zero for odd N. In a later study, Harigaya et al.[23] found a helical spin state for the Möbius strip whose total magnetization is also zero for even N. Because odd N Möbius strips have odd number of edge carbon atoms, it is not surprising that their total magnetization is non-zero. Here we focus our discussion on Möbius ribbons of even N. Since spins on one zigzag edge of graphene nanoribbons tend to line up in one direction[17] and Möbius ribbons topologically have only one edge, one would expect that Möbius ribbons should favor one spin over the other and therefore have a non-zero total magnetization. To test this hypothesis, we use spin-polarized density functional theory to examine spin states of zigzag-edged Möbius graphene nanoribbons with different widths and lengths in the present work. We will show that they indeed have a ground state with non-zero total magnetization.

## 2. Method

The Vienna Ab Initio Simulation Package (VASP)[24,25] was used to perform DFT calculations with planewave bases and periodic boundary conditions and within the generalized-gradient approximation (GGA) for electron exchange and correlation.[26] Projector-augmented wave (PAW) method[27,28] was used within the frozen core approximation to describe the electron-core interaction. A kinetic energy cutoff (450 eV) was used. A Möbius ribbon was put in a large simulation box (20 Å × 20 Å × 25 Å). Only the Γ-point was used for k-sampling. All atoms in the unit cell were allowed to relax and the force tolerance was set at 0.025 eV/Å. Local magnetic moments were obtained by integrating the site-projected density of states for both spin up and spin down electrons up to the Fermi level and then taking the difference.

## 3. Results and discussion

A zigzag-edged Möbius graphene nanoribbon can be defined by a width (N) and a length (L), as shown in Figure 1. The narrowest zigzag-edged Möbius graphene nanoribbon is Möbius cyclacene (that is, N=2). Previously, Möbius cyclacene has been studied with semi-empirical and *ab initio* methods, but the spin degree of freedom was not included in those studies.[29,30] We show spin-polarized DFT-GGA results for various spin states for cyclacene and Möbius cyclacene with a length of L=15 in Table 1. One can see that the open-shell singlet is the most stable for the cyclacene and the triplet's energy is higher than the closed-shell singlet state. However, the triplet becomes the ground state for the Möbius cyclacene, followed by the open-shell singlet. Higher spin states are less stable than the closed-shell singlet for both cyclacene and Möbius cyclacene. Figure 2 shows spin magnetization density for the open-shell singlet and triplet states of the Möbius cyclacene. For the open-shell singlet, one can see that spin up and spin down electrons are divided into two domains around the ribbon and the domain wall is near the top of the figure. At the bottom of the figure, part of the ribbon has spin up at one edge and spin down across the ribbon. For the triplet, one can see that spin up electrons dominate the edge and minor down spins appear at the inner edge (at the upper part of Figure 2b, where there is a significant twist in the ribbon).

To examine the effect of ribbon width, we examined zigzag-edged Möbius graphene nanoribbons with N=4. Table 2 shows the energetics for various spin states with different lengths. Comparing Möbius cyclacene with zigzag-edged Möbius graphene nanoribbon [4,15], one can see that the latter also has a triplet ground state but its higher spin states are also lower in energy than its open-shell singlet. The reason for this increased stability for higher spin states of wider ribbons can be traced to flat zigzag-edged graphene nanoribbons. We found that polyacene (that is, N=2) does not have a ferromagnetic metastable state between the nonmagnetic and antiferromagnetic states,[20] while wider (N>2), infinitely long zigzag-edged graphene nanoribbons do have an ferromagnetic metastable state close in energy to the antiferromagnetic state.[18] We also found higher-spin metastable states for finite-sized zigzag-edged graphene nanoribbons of increased length.[19] When wider (N>2), finite-sized zigzag-edged graphene nanoribbons are transformed into zigzag-edged Möbius graphene nanoribbons, the higher-spin states become either the ground state or low-lying metastable states. This point is further supported by data in Table 2: for zigzag-edged Möbius graphene nanoribbon [4,17] and [4,18], the ground state becomes the quintet, indicating a preference for higher total magnetization.

Another interesting result from Table 2 is that for zigzag-edged Möbius graphene nanoribbon [4,18], the open-shell singlet's energy is higher than that of the triplet and close to that of the septet. This prompted us to consider the limit of ribbon length going to infinity. When the Möbius ribbon is long enough, it should approach the corresponding zigzag-edged graphene nanoribbon in its properties, which means that the open-shell singlet (corresponding to the antiferromagnetic state for zigzag-edged graphene nanoribbons) should become the ground state. Although the lengths in Table 2 have not reached the crossover yet, data for zigzag-edged Möbius graphene nanoribbon [4,18] indicate that the open-shell singlet's energy is already higher than the triplet state.

We plot the spin magnetization density for the quintet of zigzag-edged Möbius graphene nanoribbon [4,18] in Figure 3. It can be seen that both spin up (the majority) and spin down (the minority) electrons are mainly localized at the edge. Figure 4 shows the local magnetic moments along the zigzag edge. One can see that spins on the topologically single zigzag edge are divided into four magnetic domains: two spin up and two spin down. The two same-spin domains are also different in size. At the domain boundaries, spins are frustrated, due to the twist and the connectivity induced by the Möbius boundary condition.

In a previous study employing the Hubbard model within the unrestricted Hartree-Fock approximation, Wakabayashi and Harigaya[22] found that the total magnetizations for zigzag-edged Möbius graphene nanoribbons [4,10] and [4,20] are both zero. Our first-principles results are in contrast with theirs. We have shown that zigzag-edged Möbius graphene nanoribbons with N=2 and N=4 all have a ground state of non-zero total magnetization. Our finding verifies our hypothesis that zigzag-edged Möbius graphene nanoribbons favor one spin over the other along the edge because the Möbius topology leads to one zigzag edge and spins on the same zigzag edge tend to line up. In a later study, Harigaya et al.[23] found a helical spin state for zigzag-edged Möbius graphene nanoribbon [4,10] with zero total magnetization as well. In the present work, we have not considered non-collinear spins for zigzag-edged Möbius graphene nanoribbons from first principles. We plan to do that in the future, to further test our hypothesis of a non-zero total magnetization for zigzag-edged Möbius graphene nanoribbons.

Given recent discussions about reliability of DFT methods in hydrocarbons,[31-33] it is worthwhile to discuss the applicability of spin-polarized (that is, unrestricted) DFT-GGA method employed in this work to zigzag-edged Möbius graphene nanoribbons examined here. First, the large errors of DFT methods in evaluating hydrocarbon isomer energy differences and formation energies and the lack of dispersion interactions in DFT methods as discussed recently[31-33] are not a major concern here because we are comparing different spin states for the same structure; in other words, the dispersion interactions would be equal for all spin states and cancel each other when we compare their relative energetics.

Second, spin-polarized DFT methods are generally a very affordable tool for modeling magnetic systems (molecules and solids) from first principles. However, caution must be exercised when studying spin states of transition-metal compounds.[34] Moreover, open-shell singlet systems such as some diradicals require multi-reference methods.[35,36] But spin-polarized DFT-GGA seems to describe the antiferromagnetic singlet state of zigzag-edged graphene nanoribbons satisfactorily. For example, we have applied spin-polarized DFT-GGA to acenes previously[20] and found that higher acenes have an antiferromagnetic singlet ground state. Chan and coworkers[37] have reached the same conclusion by performing complete active space calculations with *ab initio* density matrix renormalization group algorithm and treating correlation fully in the π-valence space, a high-level wave function and computationally demanding treatment. Third, spin contamination (that is, the difference between $<S^2>$ and $<S_z>$) is allowed for DFT methods because the single determinant of Kohn-Sham orbitals is the wave function of the noninteracting system, not the real system.[38] Spin contamination is usually less for DFT methods than for the unrestricted Hartree-Fock method.[39] Although the specific program we used in this work does not evaluate $<S^2>$, we expect that spin contamination would be small for spin states of nanoribbons examined here.[40]

## 4. Summary and conclusions

We have employed spin-polarized density functional theory within the generalized-gradient approximation to examine spin states for zigzag-edged Möbius graphene nanoribbons with different widths and lengths. For Möbius cyclacene, we found that the ground state is a triplet and the open-shell singlet is the next stable state, in contrast with the cyclacene counterpart. For wider zigzag-edged Möbius graphene nanoribbons, we found that the total magnetization of the ground state increases with the ribbon length. For example, zigzag-edged Möbius graphene nanoribbon [4,15] has been found to have a triplet ground state, while [4,18] has a quintet ground state. Zigzag-edged Möbius graphene nanoribbons' preference for a ground state of non-zero total magnetization can be attributed to the fact that they topologically have only one zigzag edge and the bipartite lattice structure that gives flat zigzag-edged graphene ribbons the antiferromagnetic ground state is lost in zigzag-edged Möbius graphene nanoribbons. Analysis of spin magnetization density showed that spins are mainly localized at the edge of zigzag-edged Möbius graphene nanoribbons and form domains along the edge.

**Acknowledgement.** This work was supported by Office of Basic Energy Sciences, U.S. Department of Energy under Contract No. DE-AC05-00OR22725 with UT-Battelle, LLC, and used resources of the National Center for Computational Sciences at Oak Ridge National Laboratory and the National Energy Research Scientific Computing Center, which is supported by the Office of Science of the U.S. Department of Energy under Contract No. DE-AC02-05CH11231.

Table 1. Relative energies (with respect to the closed-shell singlet, in eV) of cyclacene and Möbius cyclacene with different total magnetization (M, in $\mu_B$). (Ribbon length L=15)

| Spin state | Open-shell singlet | Triplet | Quintet | Septet |
|---|---|---|---|---|
| M | 0 | 1 | 2 | 3 |
| Cyclacene | -0.21 | 0.07 | 0.10 | 1.14 |
| Möbius Cyclacene | -0.11 | -0.17 | 0.18 | 0.81 |

Table 2. Relative energies (with respect to the closed-shell singlet, in eV) of zigzag-edged Möbius graphene nanoribbons with different total magnetization (M, in μ$_B$) and length (L). (Ribbon width N=4)

| Spin states | Open-shell Singlet | Triplet | Quintet | Septet |
|---|---|---|---|---|
| M | 0 | 1 | 2 | 3 |
| L=15 | -0.53 | -0.67 | -0.61 | -0.63 |
| L=16 | -0.32 | -0.42 | -0.37 | -0.35 |
| L=17 | -0.61 | -0.63 | -0.73 | -0.71 |
| L=18 | -0.38 | -0.27 | -0.55 | -0.41 |

Figure captions:

Figure 1. Formation of a zigzag-edged Möbius graphene nanoribbon from a zigzag-edged nanographene. The Möbius ribbon is formed by connecting A to A', B to B', C to C', and D to D'. The width (N=4) and length (L=18) defines zigzag-edged Möbius graphene nanoribbon [4,18]. Carbon atoms in black and H atoms in light gray.

Figure 2. Isosurfaces of spin magnetization density (ρ↑ − ρ↓) for (a) the open-shell singlet and (b) triplet states of zigzag-edged Möbius graphene nanoribbon [2,15], a Möbius cyclacene. Dark and light isosurfaces are 0.05 and -0.05 e/Å$^3$, respectively.

Figure 3. Isosurfaces of spin magnetization density (ρ↑ − ρ↓) for the quintet state of zigzag-edged Möbius graphene nanoribbon [4,18]. Dark and light isosurfaces are 0.075 and -0.075 e/Å$^3$, respectively

Figure 4. Local magnetic moments of zigzag carbon atoms along the edge of zigzag-edged Möbius graphene nanoribbon [4,18] for the quintet state. How we index the edge carbon atoms (that is, those connected to hydrogen; see Figure 1) is shown in Figure 3.

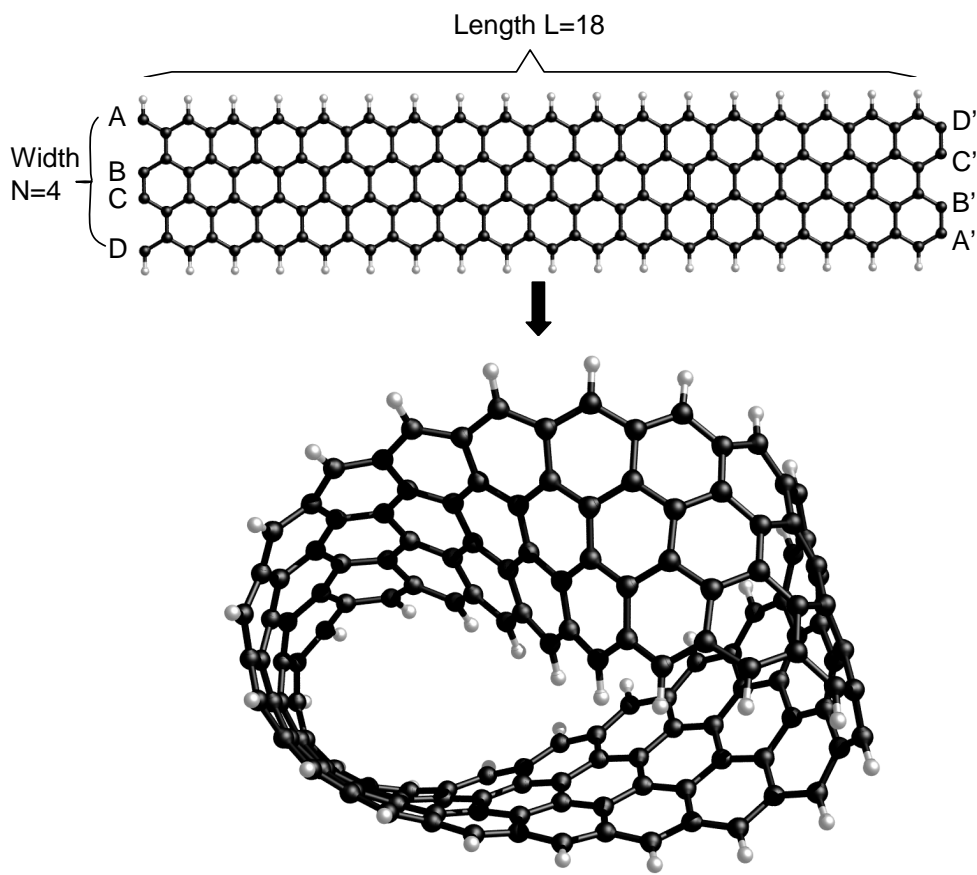

Figure 1.

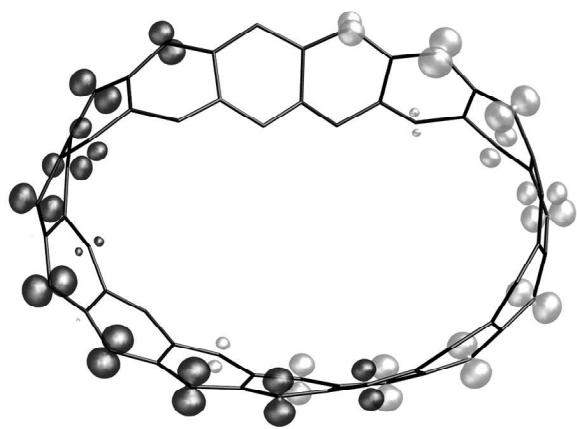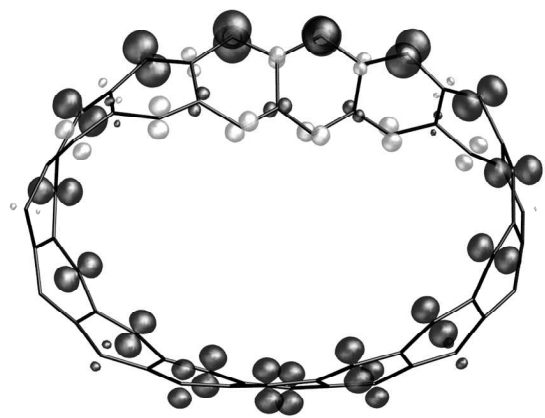

Figure 2.

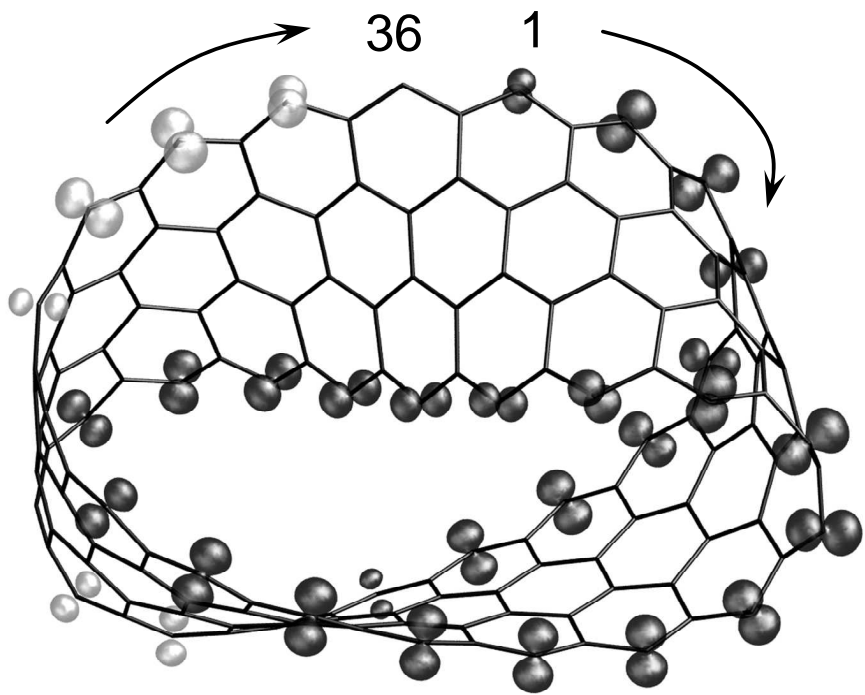

Figure 3.

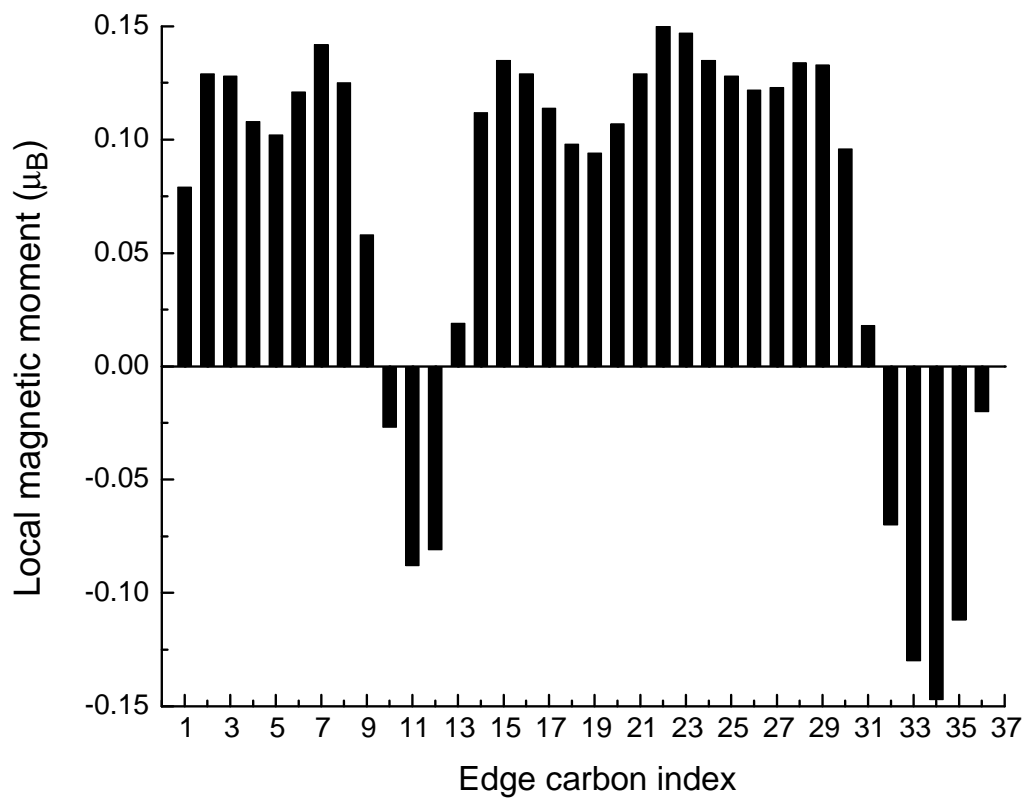

Figure 4.